\documentclass[aps,prb,showpacs,floatfix,amsmath,amssymb,superscriptaddress,preprint]{revtex4-1}
\usepackage{color}
\usepackage{graphicx}
\usepackage{natbib}
\usepackage{epsfig}
\usepackage{setspace}
\usepackage{amsmath,amssymb}
\usepackage{verbatim}
\usepackage{bibentry}

\begin{document}
\title{Small Heterocyclic Molecule as Multistate Transistor : A Quantum Many-body Approach}
\author{Dibyajyoti Ghosh}
\affiliation{Chemistry and Physics of Materials Unit, JNCASR,
Bangalore-560064, Karnataka, India}
\author{Prakash Parida}
\affiliation{Department of Physical Sciences,,
Central University of Punjab, Bathinda, Punjab}
\author{Swapan K. Pati}\email{pati@jncasr.ac.in}
\affiliation{Theoretical Sciences Unit, JNCASR,
Bangalore-560064, Karnataka, India}

\begin{abstract}
\noindent

{Weakly coupled molecular junctions are quite active and important field of 
research as they exhibit various non-linear transport phenomena. We have 
investigated the carrier transport through weakly coupled
$\mathrm{B_{2}C_{2}N_{2}H_{6}}$ molecules using quantum many body approach 
coupled with kinetic (master) equations. Various types of non-linear current 
voltage characteristics, such as, negative differential conductance (NDC), 
rectifications, Coulomb staircase, which are hallmark of multi-state transport 
devices, have been obtained. Source-drain voltage induces changes in the 
occupation probabilities of low-lying many body states depending on the 
nature of the carrier transport. The nature of the carrier transport directly 
controls the net current flowing through the molecular junctions. We further 
investigated the effect of different kinds of perturbations such as 
application of gate voltage and magnetic field perpendicular to the  
direction of the molecule upon the carrier-flow through various molecular 
bridges. Interestingly, we find that depending on the strength of the 
applied perturbing field, several phenomena, such as, switching off of the
current, suppression of NDC etc. appears in the device. Fundamentally, this 
applied perturbation modifies both the site charge density as well as 
occupation probabilities of transport active channels, resulting in 
significant alteration in the transport behavior of the molecular junction. 

}
\end{abstract}
\pacs{}

\maketitle
\section{Introduction}

Molecular electronics, the investigation of the electronic properties of 
circuits which are composed of individual molecules, have gained a huge 
research attention in past the few 
decades \cite{heath2003molecular,cuniberti2006introducing}. Appearance of 
efficient switching \cite{zhang2015towards}, negative differential conductance 
(NDC) \cite{xu2015negative}, rectification \cite{metzger2015unimolecular}, 
spin-transport\cite{rocha2005towards}, spin-filtration \cite{aravena2011coherent}, 
thermoelectric effect \cite{dubi2011colloquium} in various molecular junctions 
evidently shows the potential of these nano-devices to be used as alternative 
of conventional silicon-based semiconductor electronics. Among all these 
fascinating properties, particularly, rectification and NDC characteristics have 
been explored quite thoroughly from the very beginning of this field of 
study. Efficient rectification and / or NDC has been observed in a number 
of molecules attached strongly to the bulk electrodes. This is attributed mainly 
to (1) asymmetric nature of the molecule (donor-acceptor 
molecules) \cite{metzger2015unimolecular} (2) different electrode-molecule 
coupling strengths (asymmetric anchoring groups or electrodes)
\cite{metzger2015unimolecular} (3) different spatial potential 
profiles \cite{xu2015negative} etc. Although these phenomena mainly appear 
for covalently bonded molecule-electrode systems (i.e. coherent tunneling 
regime), recent experimental and theoretical investigations demonstrated the 
same for weakly coupled molecular junctions(sequential tunneling 
regime). \cite{park2002coulomb,xu2015negative}. Small molecules (such as 
benzene) and double quantum dots (like GaAs-based QDs) which are coupled 
weakly to metallic electrodes have shown prominent rectification and 
NDC properties, due to various factors, such as, internal charge transfer, 
intrinsic molecular asymmetries, interference effects, Pauli spin-blocking 
etc. \cite{hettler2003current,parida2009negative,ono2004nuclear,xu2015negative}. 

Recently, there is a huge surge in interest in the modulation of electron 
transport through these molecular junctions by including different kinds of 
optical, magnetic and electronic perturbations. Particularly, it has been 
demonstrated that depending upon various conditions, the carrier transport 
through weakly coupled molecule-electrode systems can become very sensitive 
towards the applied external magnetic fields. The magnetic field induced 
tuning of current may arise due to several factors, such as, tuning of 
interfering electronic degenerate states, modulation of the sharp 
transmission resonances etc.  

Theoretical modeling of these weakly coupled devices show that coherent 
non-equilibrium Green’s function 
(NEGF) \cite{caroli1971direct,rocha2004asymmetric} formalism coupled with 
self-consistent field (SCF) approach is not adequate to reproduce experimental 
findings even at the qualitative level. Unlike in the case of strongly 
coupled systems, charging energies of molecules/QDs in these devices are 
much higher than electrode coupling and plays a major role (in what?) 
\cite{muralidharan2006probing,muralidharan2007generic}. 
Consequently, these devices remain in Coulomb blockade (CB) regime where 
integral charge transfer dominates the electron transport through molecular 
junctions. To describe the molecular transport in the CB regime the 
quantum master/rate equation approach is widely 
used\cite{muralidharan2007theory,hettler2002non}. This formalism efficiently 
describes electron transport through many-body eigenstates of molecular 
systems. Since charging energy is much higher than the molecule/dot electrode 
coupling, in the weak coupling limit,  we do not explicitly consider the 
electrode or its coupling with the molecule/dot device, in the kinetic equation 
method. We discuss the formalism in detail in Models and Methods section of 
the paper.

   Using this approach, Hettler et. al. have demonstrated the large NDC 
behavior in weakly coupled benzene-based molecular 
junctions \cite{hettler2003current}. They proposed that under a finite bias, 
the radiative relaxation of electrons populate a particular many-body state 
which blocks the transport of current, resulting in the NDC behavior.
Darau et al. revisited the same system with generalized master equation 
approach, where strong interference effect appears to be the reason for 
the observed NDC behavior \cite{darau2009interference}. 

Apart from molecules, donor-acceptor QDs have also been investigated 
thoroughly for their various non-linear transport characteristics in weak 
coupling regime. Muralidharan et al have demonstrated the criterion to find 
NDC in these double QDs in terms of transition rates for populating and 
depopulating the transport-active many-body 
states \cite{muralidharan2007generic}. Song et al have observed rectification 
in I-V characteristics for weakly coupled spatially separated donor-acceptor 
systems \cite{song2007molecular}. Difference in coupling strengths of these 
sites to the electrodes results in the rectification effect in these 
molecular junctions. Parida \emph{et al.} have used kinetic equation approach to 
investigate transport characteristics in donor-acceptor double QD 
systems \cite{parida2009negative}. They propose that the increased population 
of the non-conductive triplet state with increase in bias voltage and 
consequent reduction in the current transport leads to a prominent 
NDC feature.
  

In this paper, we consider a heterocyclic benzene i.e. $\mathrm{B_{2}C_{2}N_{2}H_{6}}$ (see the inset of Fig. \ref{stab_diag}) as the molecular bridge 
which is weakly coupled to the metallic electrodes on either side. 
Thus, the molecular junction is effective in Coulomb blockade regime.
Here, B, C and N sites act as acceptor, bridge and donor (from the electron 
point of view), respectively due to their intrinsic chemical nature. This 
molecule can efficiently be modeled as two identical donor-bridge-acceptor 
half-rings (B-C-N), connected to each other in end-on manner by covalent bonds. 
We construct the interacting isolated molecular Hamiltonian considering only 
the localized $\mathrm{2p_{z}}$ orbitals of B, C and N 
atoms \cite{hettler2003current,darau2009interference} since the $\sigma$ 
orbitals are at a much higher energy scale, as followed in earlier works.  
Depending upon the pair of sites involved in electrode-molecule coupling, 
we find various non-linear current-voltage characteristics, such as, 
prominent NDC, strong rectification, Coulomb staircase, in this two-terminal 
molecular device. We analyzed  the bias-dependent probability of many-body 
states by looking at the charge distributions at all atomic sites for each of 
these low-lying states (also called transport-active channels.) We further 
looked into the carrier transport in these weakly coupled systems by applying 
perpendicular magnetic field of various strength. The current transport 
appears to be quite sensitive to  the applied magnetic field.  The effect 
of magnetic field over transport strongly depends on various factors such as 
(1) strength of the magnetic field and (2) relative position of electrodes. 
The change in atomic charge distributions as well as occupation probabilities 
of low-lying states upon the application of magnetic fields hugely alters 
the current-voltage characteristics of these molecular devices. 

\section{Model and Methods}

We undertake the quantum master-equation approach to explore transport 
characteristics of $\mathrm{B_{2}C_{2}N_{2}H_{6}}$ (see the schematic in 
the inset of Fig. \ref{stab_diag} ) in the sequential tunneling i.e. Coulomb 
blockade regime. This approach formulates the carrier transport through 
a correlated system, having many-body eigenstates. In this present approach, 
the occupation probabilities of many-body states are calculated from their 
corresponding wave functions. Note that, in the present study, we neglect the
off-diagonal coherence while solving the rate equation. 

We model $\mathrm{B_{2}C_{2}N_{2}H_{6}}$ molecule by taking 6 electrons in 
6 sites within Hubbard model. Taking into account the on-site electron-electron 
interactions and hopping between nearest neighbor sites, we write down the 
most general form of the Hamiltonian as follows,

\begin{eqnarray}
H =  \sum_{i=1, \sigma}^6\ (\epsilon_i-eV_g)a^{\dag}_{i\sigma}a_{i\sigma} + 
\sum_{i\sigma } - t(a^{\dag}_{i\sigma}a_{i+1,\sigma}+h.c.)
 + U \sum_{i=1}^N n_{i\uparrow}n_{i\downarrow}  
\label{exhub}
\end{eqnarray}
\noindent
where t is hopping strength between nearest neighbor sites, $\epsilon_{i}$ 
is the on-site energy for different atomic sites, $U$ is the Hubbard interaction 
term and $V_g$ represents the external gate bias. We consider only nearest 
neighbor hopping and have taken equal hopping strength (2.4 eV) for all bonds, 
i.e., B-C, C-N and B-N \cite{hettler2003current,pati2014exact}. Here, we 
neglect the small differences among the hopping parameters which may 
appear due to different chemical nature of C, B and N atoms. For Hubbard 
parameters, we obtain Hubbard on-site energy as the difference between 1st 
ionization potential and electron affinity of each of the atomic species. Thus, 
the Hubbard on-site parameters are evaluated to be 8.02 eV, 9.67 eV and 14.46 eV 
for B, C and N, respectively \cite{sansonetti2005handbook}. 
The site-energy term in the Hamiltonian corresponds to negative of first 
ionization potential for each atomic species. And these turn out to be -8.30 eV, 
-11.26 eV and -14.53 eV for B, C and N, respectively.

To compute the transport properties in the sequential tunneling limit, we 
diagonalize the Hubbard Hamiltonian, $H$. Diagonalization of $H$ gives 
many-body eigenstates $|s>$ with the corresponding eigen energies, $E_s$. We 
compute  the occupation probabilities $P_s$, through the master equation 
approach, in the steady state of the system. The transition rate, 
$W_{s' \rightarrow s }$, from many-body state, $s'$ of the molecule with $N$ 
electrons to  a state $s$ with $(N-1)$ or $(N+1)$ electrons, is calculated 
up to linear order in $\Gamma$, where $\Gamma$ is the bare electron tunneling rate between the molecule and the left/right electrode. Using the Fermi's golden rule the transition rate can be written as follows,

\begin{eqnarray}
W_{{s^\prime}\rightarrow{s}}^{L+}=\Gamma f_L(E_s-E_s^\prime) \sum_{\sigma} 
|<s|a^\dag_{1\sigma}|s^\prime>|^2\nonumber  
\end{eqnarray}
\begin{eqnarray}
W_{{s^\prime}\rightarrow{s}}^{R+}=\Gamma f_R(E_s-E_s^\prime) \sum_{\sigma} 
|<s|a^\dag_{N\sigma}|s^\prime>|^2 
\end{eqnarray}

\noindent The corresponding equation for
$W_{{s} \rightarrow {s^\prime}}^{L-}$ and
$W_{{s} \rightarrow {s^\prime}}^{R-}$ are formulated 
by replacing $f_{L,R}(E_s-E_s^\prime)$ by
$(1-f_{L,R}(E_s-E_s^\prime))$, where $f_{L/R}$ is the Fermi function for left/right electrode. 
Here, $+/-$ represents the creation/annihilation of an electron
inside the molecule due to electron movement from/to left ($L$) or right ($R$) electrodes.
$C^\dag_{1\sigma}$ and $C^\dag_{N\sigma}$ are the creation operators for electrons with spin, $\sigma$, 
at the $1^{st}$ and $N^{th}$ lattice sites, respectively.
We also have assumed that the creation and annihilation happens only at the sites which are directly connected to the electrodes. 
The total transition rate is obtained as,
$W_{{s} \rightarrow {s^\prime}}=W_{{s} \rightarrow {s^\prime}}^{L+}+
W_{{s} \rightarrow {s^\prime}}^{R+}+W_{{s} \rightarrow {s^\prime}}^{L-}+
W_{{s} \rightarrow {s^\prime}}^{R-}$.
Now. the non-equilibrium probability, $P_s$, of occurrence of each many-body state, $s$, can be
represented by the rate equation,

\begin{eqnarray}
\dot{P_s}=\sum_{s^\prime}(W_{{s^\prime} \rightarrow s}P_{s^\prime}-W_{{s} 
\rightarrow {s^\prime}} P_s).
\label{master}
\end{eqnarray}
\noindent
At the steady state, the population of the different many-body states ($P_s$) 
can be found by solving the above rate equation ( Eq. (\ref{master}) ).
Thus, the Eq. (\ref{master}) becomes
\begin{eqnarray}
\dot{P_s}=\sum_{s^\prime}(W_{{s^\prime} \rightarrow s}P_{s^\prime}-W_{{s}\rightarrow {s^\prime}} P_s) = 0.  
\label{mastersteady}
\end{eqnarray}
\noindent
The Eq. (\ref{mastersteady}) can be written in an expanded form that results 
in a homogeneous linear system (AX=0) of the size of the many-body space.
Since AX = 0 can not be solved, we make use of $\mathrm{\sum_s{P_s}=1}$ to 
eliminate one row/column, thus reformulating the
eigenvector problem into an inhomogeneous linear system (AX=B), which
can be solved using well-known linear algebraic methods\cite{anderson1999lapack}.
Thereafter, the current in the left and right electrodes is calculated by the
following formula, 
\begin{eqnarray}
I_{\alpha} ={e \over \hbar} \sum_{s,s^\prime}(W_{{s^\prime} \rightarrow s}^{\alpha+}
P_{s^\prime}-W_{{s} \rightarrow {s^\prime}}^{\alpha-} P_s)
\end{eqnarray}
\noindent
where $\alpha=L/R$. Note that at steady state, the current at two terminals
is same, i.e. $I_{L} (t) =  I_{R} (t) = I (t)$.

Further we also study the carrier transport through these molecular junctions in 
the presence of perpendicular magnetic fields. We consider the effect of 
applied magnetic field ($B$) by modifying the hopping parameter, $t_{ij}$, 
such that 
it acquires (?)/includes(?) a Peierl's phase, as discussed in earlier 
works.\cite{wakabayashi1999electronic,rocha2005metallic} 
Thus, $t_{ij}$ modifies as $t_{ij} e^{\frac{2 \pi i \delta \phi}{\phi_0}}$, 
where $\phi_0 = h/e$ is {\underline {the quantum of magnetic flux}}, and 
 $\delta \phi$ is the Peierl's phase given by,
\begin{eqnarray}
\delta \phi = \int_i^j \textbf{A}.d\textbf{l},
\end{eqnarray}
where \textbf{A} is the vector potential created by the perpendicular 
B-field, \textbf{B} = (0, \textbar\textbf{B}x\textbar, 0). 
Magnetic flux, $\phi$, correspdoning to this magnetic field is  
$\phi$ = \textbf{B}.\textbf{$A_s$}, where \textbf{$A_s$} is the area of 
the hexagonal ring i.e. $3\surd3 a^2/2$ with a as the bond length (or radius of
the ring???).

We investigate the transport characteristics for a donor-bridge-acceptor
molecular circuit in weak-coupling regime. We simulate the molecular junction 
at room-temperature, T = 300K and consider $\Gamma$, the molecule-electrode 
coupling strength, to be 0.25 meV. We consider symmetric electrode-molecule 
coupling for the presently studied junctions. In our case, asymmetry in 
the system is inherent and appears due to the different chemical nature of 
constituent atoms. We apply the exact diagonalization 
(ED) method to diagonalize the 6-site Hamiltonian, as described in equation 
Eq. (\ref{exhub}), with a total number of basis of $6^4$ i.e. 4096. 

\underline{Several previous investigations established the fact that asymmetric coupling strengths 
for left and right electrode-molecule junction give rise to NDC or rectification in these weakly 
coupled molecular junctions \cite{hettler2002non,muralidharan2007generic}. 
However, we consider symmetric electrode-molecule coupling for the presently studied junctions. 
For our case, asymmetry in the system is inherent and appears due to very different chemical nature of constituting atoms.} \\

\section {Results and Discussion}

\begin{figure}
\centering
\includegraphics[width=0.7\textwidth]{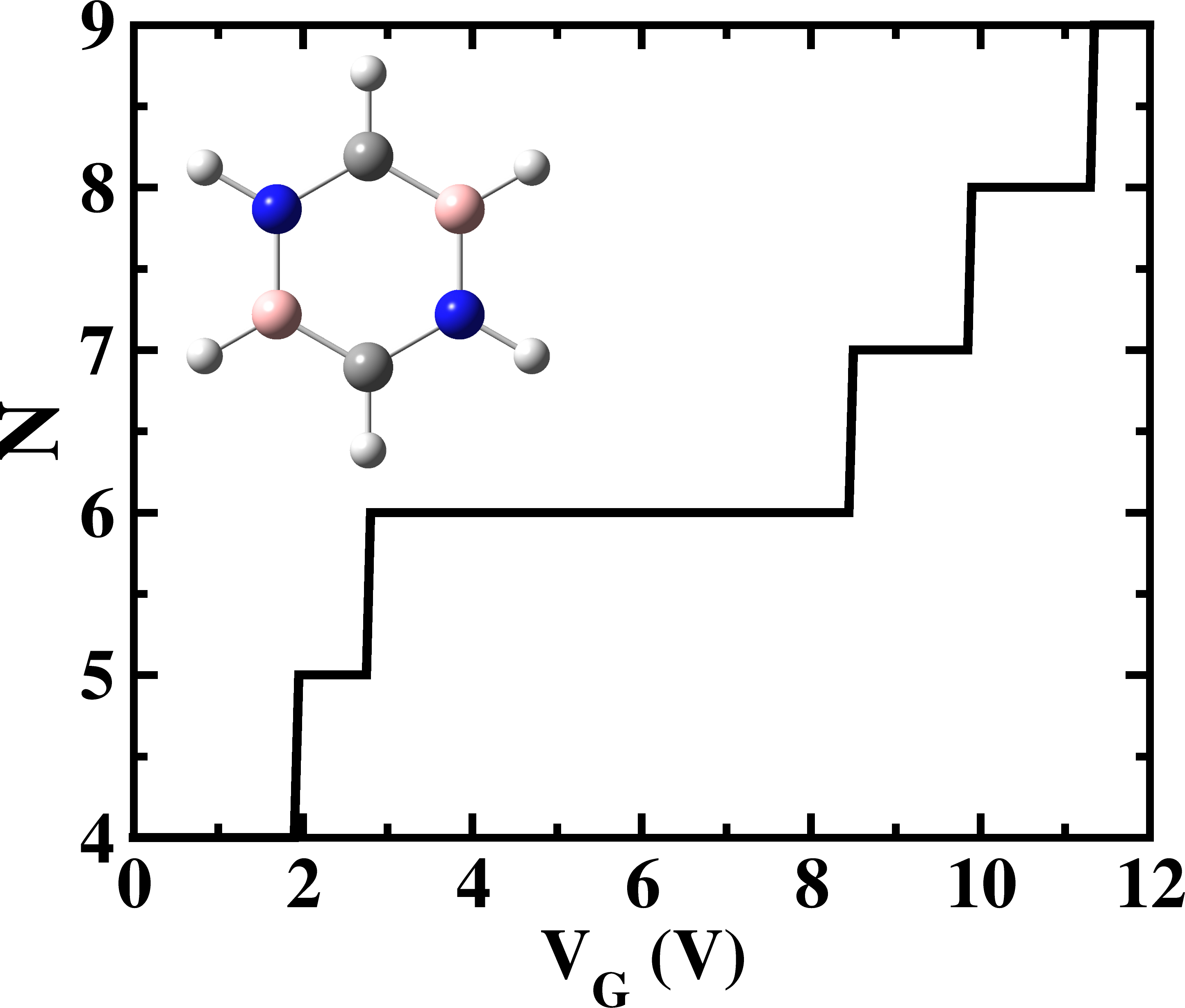}
\caption{\label{stab_diag} The number of electrons (N) in the molecule with variation of gate voltage ($V_G$) of the device.
We have not included the spin sector in the present plot. In the inset, a schematic representation of a $\mathrm{B_{2}C_{2}N_{2}H_{6}}$ molecule is given. 
Hydrogen, boron, carbon and nitrogen are represented by white, lime, gray and blue colored spheres.}
\end{figure}

We first investigate the stability of different charge states of the 
molecule under applied gate voltage ($V_G$), in the absence of magnetic field. 
We have given the number of electrons in the molecule with lowest energy under 
various gate bias values, keeping source-drain bias ($V_{SD}$) zero 
in Fig. \ref{stab_diag}. (NOT CLEAR)
As can be seen in Fig. \ref{stab_diag}, the six electron molecular state 
has widest plateau, indicating that this charge state is the ground state 
of the molecule. Consequently, we choose the Fermi energy of the device such 
that molecule remains in its ground state. 

\begin{figure}
\centering
\includegraphics[width=0.6\textwidth]{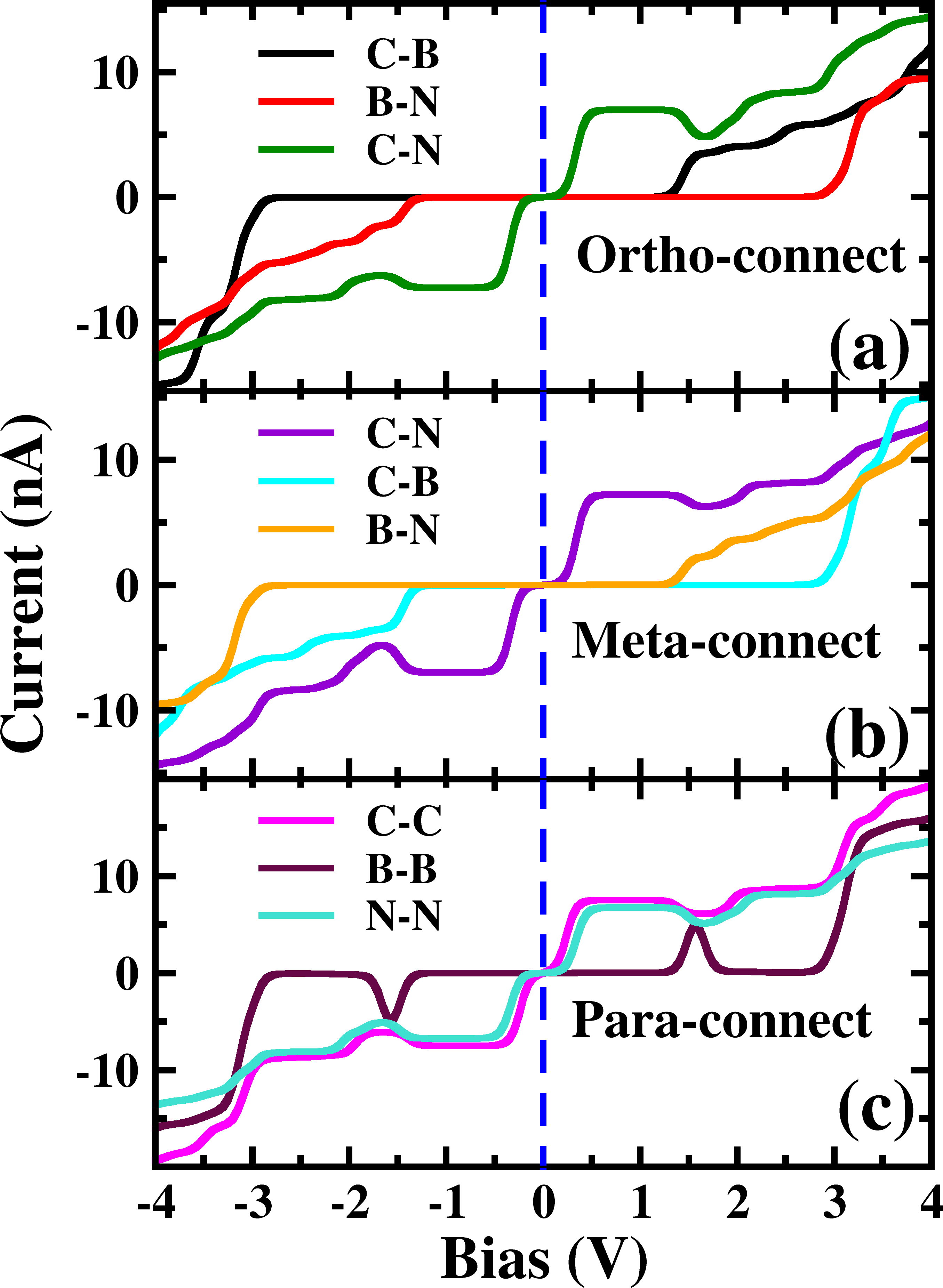}
\caption{ \label{iv_all} Current (I) - source-drain voltage ($V_{SD}$) 
characteristics of molecular device (bridge???) where electrodes are 
connected in (a) ortho, (b) meta and (c)para-position. Dashed blue line 
represents the $V_{SD}$ = 0.0. In the upper left insets of (a-c), 
the atomic sites where the electrodes are attached are shown.}
\end{figure}

As the molecule is cyclic in nature, we can connect the electrodes to it in different possible conformations such as ortho, meta or para positions. 
We further fix the Fermi energy in such a way that small $V_{SD}$ bias 
can transform the molecule from 6 electron ground state to 7 electron ground state 
i.e. anionic state with one more electron.  Note that, in this paper, we are 
investigating the I-V characteristics of the device, considering only neutral 
or anionic molecule. {CHANGE THIS SENTENCE} The I-V characteristics of three 
different connectivities are given in Fig. \ref{iv_all}. From Fig. \ref{iv_all}, 
it is evident that I-V characteristic of these junctions is highly sensitive 
towards the chemical nature of molecular site to which the electrode is 
connected. We will discuss the different kinds of  non-linear transport 
behaviours such as negative differential conductance (NDC), rectification and
staircase, in next sections.

\subsection {Negative Differential Conductance}

\begin{figure}
\centering
\includegraphics[width=0.9\textwidth]{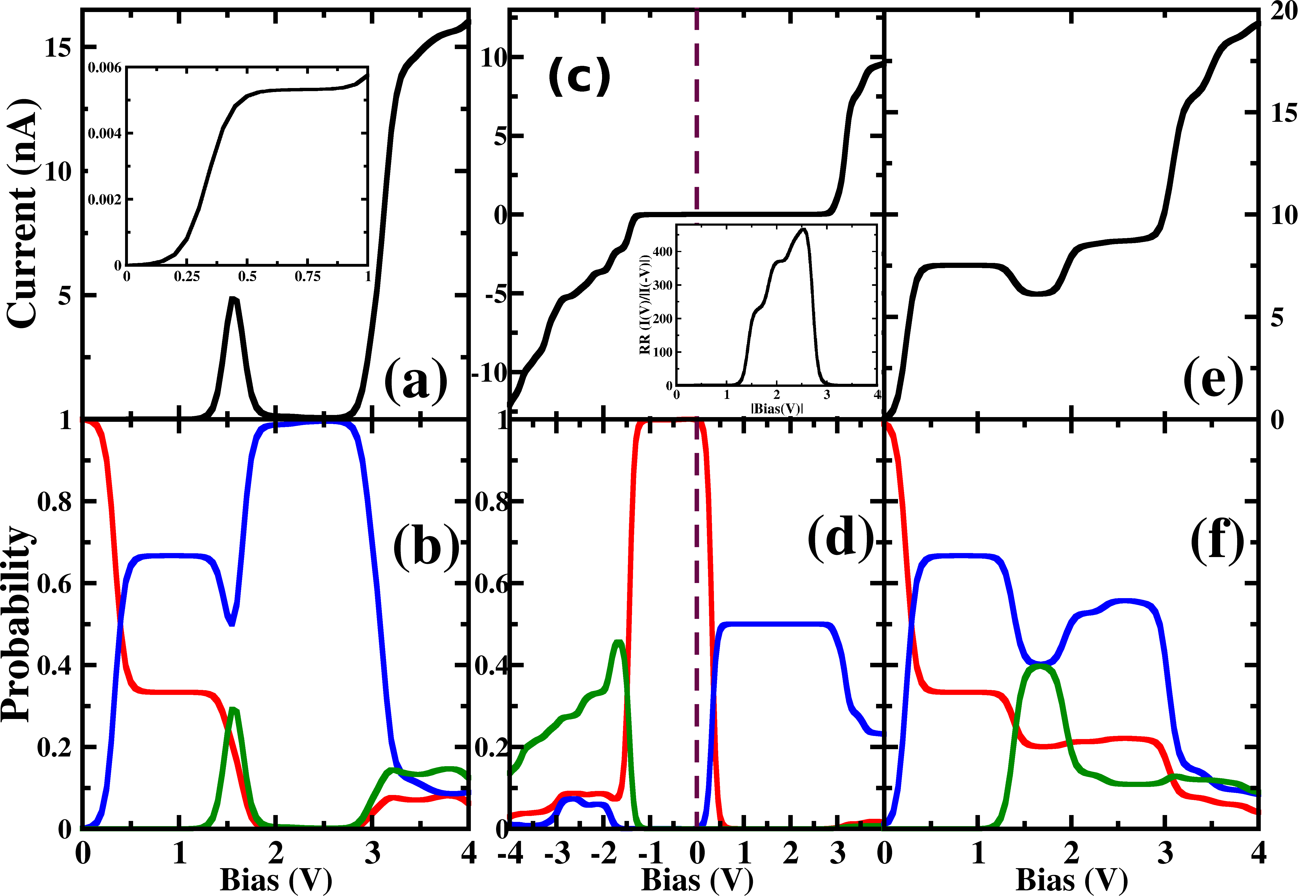}
\caption{ \label{ndcrec} The variation of current (I) with $V_{SD}$ for 
a device at different positions in benzene (top panel of the figure) 
and the occupation probabilities at the bottom panel of the figure for the
same configuration. (a) I-V for two boron atoms at para position, 
(c) I-V for a boron and nitrogen atom at ortho position and (e) I-V for two 
carbons at para position. Occupation probabilities are represented by
red line for 6e-gs, blue line for 7e-gs and green line for 7e-1st-es states 
in all the figures. In the inset of (a), I-V plot for the junction connected 
by two borons has been given for bias-range $ 0.0 < V_{SD} < 1.0 V$.}
\end{figure}

Firstly, we consider the molecular bridge, (particular device) where both 
para-positioned B atoms are weakly connected to the electrodes.
Note that, the I-V characteristic of this device is identical at both the
positive and negative bias, but with an opposite sign. i.e. I($V_{SD}$) = 
I (-$V_{SD}$).  It is quite obvious, since both the geometry and strength of
molecule-electrode coupling are symmetric in nature. This symmetric nature of 
I-V characteristic has already been observed for benzene and double quantum dot 
systems. \cite{hettler2003current,muralidharan2007generic} Taking the advantage 
of this symmetry, here we discuss the I-V characteristics, focusing only on 
positive bias regime. As shown in the inset of Fig. \ref{ndcrec} (a), at low 
bias i.e. $V_{SD} < 1.0 V$, the current appears to be very small, of the order 
of pA only. However, most important feature of this nano-junction appears 
at the bias range of $1.0 < V_{SD} < 2.2 V$, where current rises sharply and 
then suddenly decreases with the increase in $V_{SD}$, exhibiting a strong 
NDC behavior (see Fig. \ref{ndcrec} (a)). With further increase in bias, 
i.e. $V_{SD} > 2.80 V$, the transport current steeply increases. 


To find the reason behind the appearance of NDC in this molecular junction, 
we focus on the non-equilibrium occupation probabilities of transport-active 
many-body states and analyze how they get modified  with the change in 
$V_{SD}$. As shown in Fig. \ref{ndcrec} (b), at $ 0.0 < V_{SD} < 0.2 V$, 
the 6 electron ground-state (i.e. 6e-gs) remains almost completely populated, 
leaving all other states nearly empty. In this bias region, the applied 
$V_{SD}$ is not enough to charge the molecule with an extra electron i.e. 
formation of anion is not favored. As the molecule is stable in neutral 
6 electron state and there is no other eigenstate(s) within this energy 
range, there is no current through the molecular bridge. Consequently, the 
device remains in Coulomb blockade regime. As $V_{SD}$ increases, 7 electron 
ground-state (i.e. 7e-gs) becomes energetically accessible and there is
transition of electrons from 6e-gs to 7e-gs, resulting in an increase in 
current (see Fig. \ref{ndcrec} (a) and (b)). However, unlike other molecules 
explored previously \cite{hettler2003current}, the current due to this 
particular transition, appears to be very small, of the order of a few pA, 
as mentioned earlier. It simply indicates that 7e-gs is not suitable for 
electronic transport through the present molecular device. With further 
increase in $V_{SD}$ above a certain threshold bias-voltage, 
i.e. $V_{SD} > 1.15 V$, 7 electron first excited state (7e-1st-es) becomes 
accessible. Consequently, the corresponding occupation probability starts 
increasing and exhibits sharp increase in the current for $V_{SD} > 1.30 V$ 
(see Fig. \ref{ndcrec} (a)). Importantly, here, the magnitude of current 
is of the order of nA, much higher than the current due to the {\bf tunneling}
from 6e-gs to 7e-gs. Interestingly, at $V_{SD} > 1.55 V$ the probability of 
this conducting 7e-1st-es reduces, causing sharp drop in the current. In 
this particular bias-voltage, fundamentally, the transition from 7e-1st-es 
to 6 electron first excited states (6e-1st-es) becomes energetically possible. 
Thus, the net occupation probability of 7e-1st-es starts reducing and 
consequently, for 6e-1st-es probability should increase. However, as the 
transition from 6e-1st-es to 7e-gs is also probable at the same bias range, 
the net occupation probability of 7e-gs starts increasing as can be clearly 
seen in Fig. \ref{ndcrec} (b). As 7e-gs are weakly conducting in nature, the 
increase in probability of this state results in the reduction of transporting 
current, exhibiting the NDC.  

\begin{figure}
\centering
\includegraphics[width=0.7\textwidth]{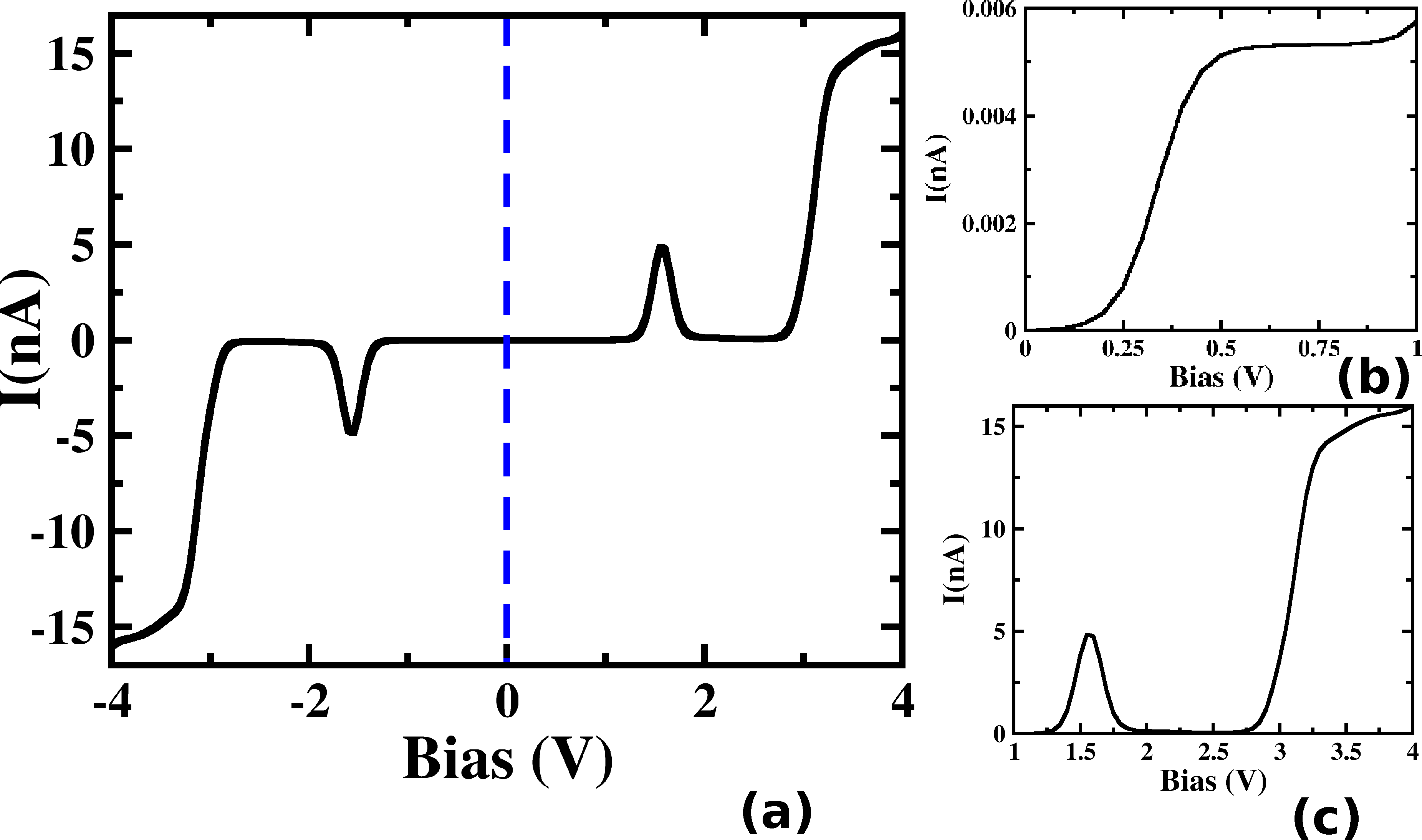}
\caption{ \label{charge_den} The charge density distribution of the 6 electron ground state, 7 electron ground state and 7 electron 1st excited state 
over the sites of $\mathrm{B_{2}C_{2}N_{2}H_{6}}$.
{\bf WRONG FIGURE} }
\end{figure}

To understand the atomistic reason behind the different nature of 7 electron 
many-body states towards the electron transport in the sequential tunneling 
regime, we have analyzed the charge densities of these states. In Fig. 
\ref{charge_den}, we represent the charge density of every sites of the 
molecule at their different low-energy transport active many-body 
states. As shown in Fig. \ref{charge_den}, quite obviously, boron and nitrogen 
atoms in 6e-gs are electron deficient and rich, respectively, due to their 
intrinsic electronegativity. Interestingly, 7e-gs also has similar  
electron density distribution where the state remains largely polar due to 
charge localization at nitrogen and charge depletion at boron sites. Hence, 
when the electrodes are connected to two electron deficient boron atoms at 
para-position, transition of molecule from 6e-gs to 7e-gs at finite bias does 
not raise the electron-density at the electrode-molecule coupling region. 
And consequently, it results in very weak current flow through molecular 
junction. This explains the negligible current flow in the bias range of 
$ 0.0 < V_{SD} < 1.3 V$ and $ 1.85 < V_{SD} < 2.75 V$ in the junction as 
shown in Fig. \ref{ndcrec} (a). On the other hand, in the 7e-1st-es, electrons 
remain almost equally distributed over all the atomic sites. Thus, when the 
7e-1st-es gets populated at particular bias, the electron densities at B atoms 
increase appreciably.  And consequently, the current of magnitude of few nA, 
flows through the molecular junction. This conduction of electron gets 
quenched when the higher bias (i.e.$ V_{SD} > 1.55 V$) populates the charge 
localized 7e-gs once again. At much higher bias i.e. $V_{SD} > 2.80 V$ many 
more excited states start appearing as transport-active channels and
consequently, the rapid growth of current can clearly be found in 
Fig. \ref{ndcrec}(a). 

Thus, the population and depopulation of anionic ground state and 1st excited 
state of $\mathrm{B_{2}C_{2}N_{2}H_{6}}$ where charge distribution patterns 
are quite different and decisive for current flow, results in exciting 
non-linear behavior of current-voltage characteristics.

\subsection{Rectification}

Considering the device where electrodes are connected to chemically asymmetric 
ortho positioned B and N atoms, a prominent rectification of current appears 
in the I-V characteristics, as plotted in Fig. \ref{ndcrec} (c). Although at 
positive bias regime, current  of magnitude of a few nA flows through the 
device, the negative bias regime shows the appearance of nA current only 
after $V_{SD} = -2.80 V$. To have a quantitative measure about rectification, 
we further calculate rectification ratio (RR) which is defined as the ratio 
between the absolute current-values at positive and negative voltage, 

\begin{eqnarray}
RR (V) = \mid I(V) / I(- V) \mid 
\end{eqnarray}

We have plotted the rectification ratio in the inset of Fig. \ref{ndcrec} (c). 
As can be seen, the RR(V) is as high as 466.2 at 2.55 V. The occupation 
probability analysis of many-body states shows that at negative bias, major 
transition occurs between 6e-gs to 7e-1st-es, resulting in the flow of current 
at $V_{SD} < - 1.25 V$ (see Fig. \ref{ndcrec} (c, d)). In this negative bias 
region, the 7e-gs remain almost empty. On the other hand, on applying a 
positive bias of $V_{SD} > 0.05 V$ to this molecular junction, occupation 
probability decreases for 6e-gs and increases for 7e-gs. However, the current 
remains negligible until $V_{SD} > 2.80 V$. After this threshold voltage, 
current starts flowing due to the increase of occupation probability of 
energetically higher lying conducting states. 

To find the fundamental reason for different conducting nature of these 
many-body anionic states, we look back at the charge density distribution per 
sites, as shown in Fig. \ref{charge_den}. It is evident that the boron site, 
attached with one of the electrode, is electron deficient for 7e-gs. Thus, 
the occupation probability of 7e-gs eventually gives rise to unfavorable 
electron transport through this molecular junction. However, for the 
7e-1st-es, both the connecting sites i.e. N and B are electron rich in nature 
and consequently appears to be suitable for electron transport. So, different 
nature of occupation probability of these low-lying states at positive and 
negative bias, results in prominent rectification in the device. Note that, 
this kind of rectification also appears for other connectivities, where the 
atomic sites, connected to the electrodes, are chemically different, as can 
be seen in Fig. \ref{iv_all}. 

\subsection{Staircase}

In the following section, we focus on the molecular junctions where electrodes 
are connected to two carbon atoms which are in para-position to each other. 
As shown in Fig. \ref{ndcrec} (e), the I-V characteristic of this particular 
device shows a step-like feature which is quite common for the junctions 
where sequential tunneling is the major mechanism for electron transport.
As discussed previously, in this sequential tunneling regime, the charging 
energy for molecules is quite high and needs to be overcome by applying a 
certain $V_{SD}$. Before this threshold $V_{SD}$, we find Coulomb blockade 
regime in the device (see Fig. \ref{ndcrec} (e)). For the present case, since 
we had chosen the Fermi level to be very close to the transition of 
6 electron state to 7 electron state, the Coulomb blockade regime appears to 
be small. Further, as shown in Fig. \ref{ndcrec} (f), with increment in 
$V_{SD}$, 7e-gs starts filling up and electrons start flowing through the 
junction. With increment in $V_{SD}$, other excited states of different charge 
and spin sectors start appearing in the active bias-window, which results in 
steps in the I-V plot for the molecular junction. 

Importantly, for this particular device, as the electrodes are connected 
to two carbon atoms which are electron-rich in nature for 7e-gs, we find 
that electron flows through the molecular junctions, as increased 
$V_{SD}$ populates this anionic state. Small dip in the current for the 
bias range of $ 1.1 V< V_{SD} < 2.0 V$ appears due to the finite occupation 
probability of 7e-1st-es which is less favorable towards electron transport 
in the present device. Lesser electron density at the carbon sites connected
to electrodes, reduces the conducting nature of 7e-1st-es than that of 7e-gs. 

Further, as shown in Fig. \ref{iv_all} (c), device with electrodes connected to para-positioned nitrogen atoms, also shows same kind of I-V characteristics.  

\subsection{Effect of Magnetic Field}

\begin{figure}
\centering
\includegraphics[width=0.7\textwidth]{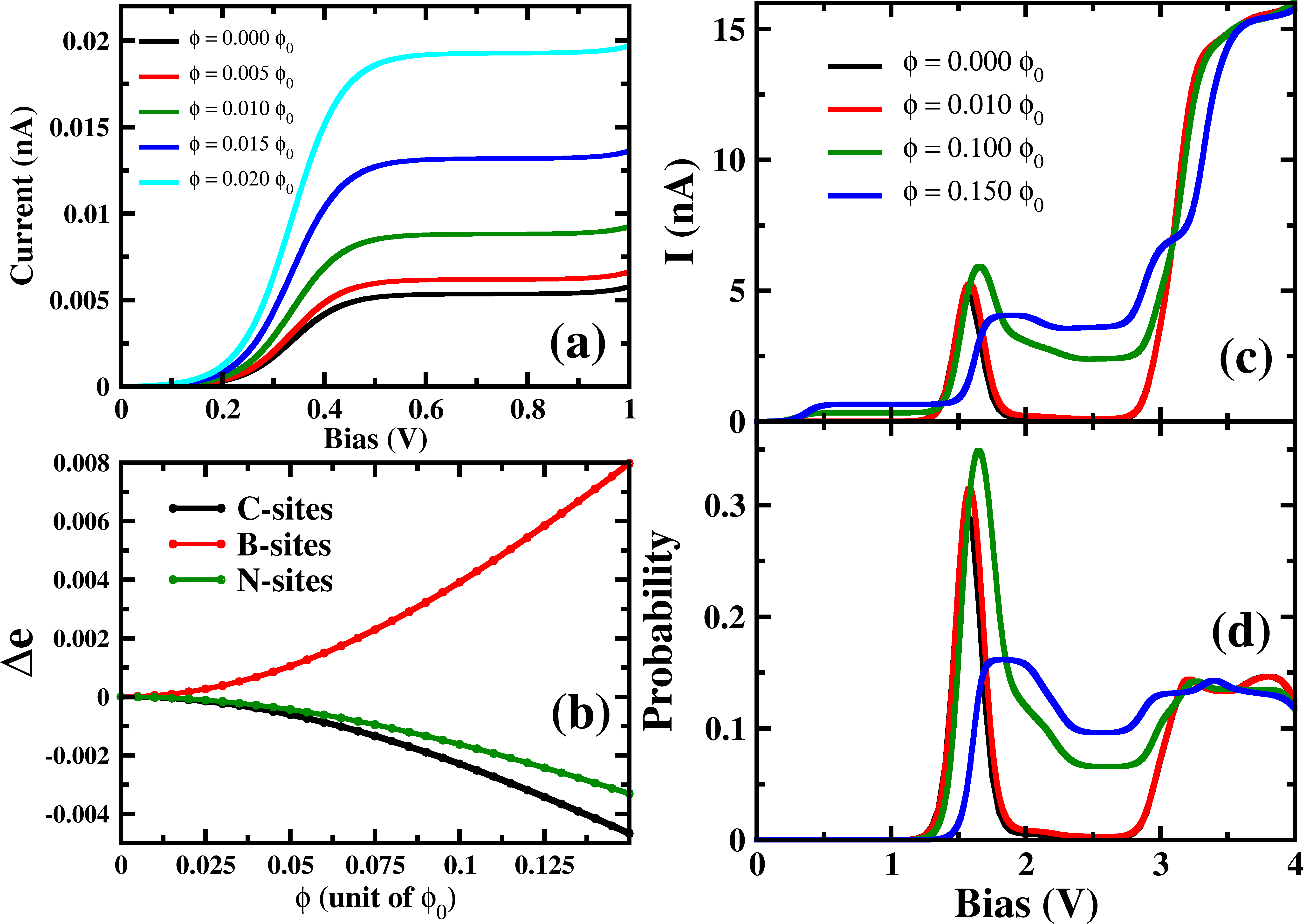}
\caption{ \label{iv_mag_1} (a) I-V characteristics for the low bias regime
 ($V_{SD}$ of $ 0 V< V_{SD} < 1.0 V$) versus magnetic field strength for 
para positioned boron connected device. (b) Charge density at boron sites 
with varying magnetic field strength. (c) I-V characteristics and 
(d) Occupation probabilities of 7e-1st-es for higher $V_{SD}$,
 for varying magnetic field strength.}
\end{figure}

\begin{figure}
\centering
\includegraphics[width=0.7\textwidth]{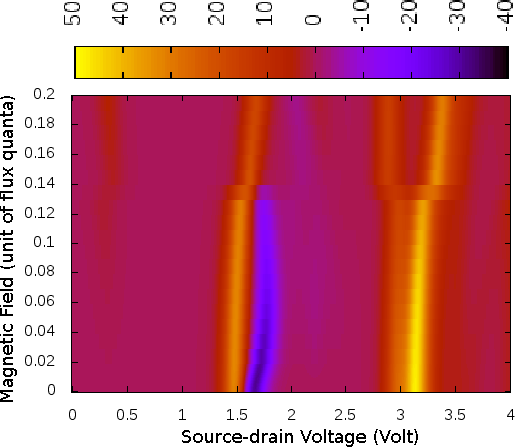}
\caption{ \label{iv_mag_2} The color map shows differential conductance (dI/dV) as a function of $V_{SD}$ and applied perpendicular magnetic field ($\phi$) for
para positioned boron connected device.}
\end{figure}

In this section, we focus on the modulation of conductance of above mentioned 
molecular junction by applying a perpendicular magnetic field. The effect of 
the magnetic field with a wide range of strength has been investigated in this 
study. Firstly, we focus on the conformation where electrodes are connected 
to two B atoms at para-position. As shown in  Fig. \ref{iv_mag_1} (a), at 
relatively small magnetic field strength i.e. in the range of 
$0.001{\phi_0} < \phi < 0.02{\phi_0}$, the current increases at low bias 
($ 0.15 V < V_{SD} < 0.50 V$) region with applied magnetic field strength. 
However, it is to be noted that the magnitude of resulting current remains in 
the pA order. The prominent NDC also remains almost unaltered for application 
of these relatively small magnetic fields. Interestingly, as we increase the 
magnetic field strength further, i.e. $0.02{\phi_0} < \phi < 0.10{\phi_0}$ 
nature of current conduction starts changing at the NDC-peak region i.e. 
$ 1.0 V < V_{SD} < 3.0 V$ (see Fig. \ref{iv_mag_1} (c)). As shown in  
Fig. \ref{iv_mag_2} (a) {\bf or (c)?????}, applying magnetic field of 
$\phi > 0.14{\phi_0}$, the NDC-peak almost disappears and I-V exhibits 
staircase behavior. Thus, depending on the strength of magnetic field, 
we can modulate the transport characteristics of this molecular junction quite 
drastically. 

To understand the modulation in I-V characteristics, we now concentrate on 
two mainly sensitive factors which get modified with applied magnetic fields- 
(1) the change in charge distribution on atomic-sites of the molecule and 
(2) occupation probabilities of transport-relevant six and seven 
electron many-body states.

On applying  a relatively small magnetic field i.e. $0.001{\phi_0} < 
\phi < 0.02{\phi_0}$, though the occupation probabilities of low-lying 6e and 
7e-states remain almost unaltered, the charge distribution on the atomic sites 
get modified for them. Most importantly, as shown in  Fig. \ref{iv_mag_1} (b), 
B-sites which are highly electron deficient in the absence of magnetic field 
in 7e-gs hinders the current conduction and start getting populated with 
electrons with the application of the magnetic field. Charge density at 
B-sites for 7e-gs keeps on increasing as we increase the applied magnetic 
field strength. This increase of electron density on B-sites in turn makes 
7e-gs more and more suitable for current conduction. Consequently, at 
low-bias i.e. $ 0.15 V< V_{SD} < 0.50 V$ where 6e-gs to 7e-gs transition 
occurs, amplitude of resulting current keeps on increasing with applied 
magnetic field strength. However, note that, at relatively small magnetic 
field, the increment of charge on B-atoms are small i.e. in the range of 
$2\times10^{-4}e^{-}$ and the resulting current at low-bias remains in 
pA only (see Fig. \ref{iv_mag_1}). Thus, it is quite evident now that current 
through boron-connected molecular junction increases along with 
the accumulation of charge-densities of boron-sites. 

As shown in Fig. \ref{iv_mag_1} (c), with higher applied magnetic field i.e. 
$0.02{\phi_0}< \phi < 0.13{\phi_0}$, where the prominent NDC peak gets 
affected, current does not completely switch-off at $V_{SD} > 1.80 V$.  
Here, we look into the atomic charge densities as well as occupation 
probabilities of relevant low-lying many-body states. As the B-sites of 
7e-gs keeps on getting populated with application of stronger magnetic 
field, the current increases at the bias range of $0.15 V < V_{SD} < 0.50 V$ 
as discussed previously. More interestingly, it is apparent from 
Fig \ref{iv_mag_1} (d) that though the occupation probability of highly 
conducting 7e-1st-es at $V_{SD} > 1.55 V$ reduces, unlike previously, it 
remains finite at the bias range, $V_{SD} > 1.80 V$. As this highly conducting 
state remains populated at the higher bias, the current never drops down 
to zero at any higher bias. Moreover, since 7e-gs is also quite conducting in 
nature, high occupation of this state also results in a measurable current 
flow through the molecular junction. Thus, at higher magnetic field strength, 
both charge densities at atomic sites as well as occupation probabilities of 
current-conducting states get modulated, resulting in change in the 
I-V characteristics. 

When we apply higher magnetic field i.e. $\phi > 0.14{\phi_0}$, as shown in Fig \ref{iv_mag_2}, the NDC disappears completely and shows staircase behavior. 
Under these magnetic fields, at low-bias i.e. $V_{SD} > 0.20 V$ the current 
gets switched-on and it keeps on increasing as the bias-voltage increases in 
a stepwise manner. From occupation probability, it is clear that current flow
at low-bias is assisted by 6e-gs to 7e-gs transition, as mentioned earlier. 
Importantly, as the 7e-gs is now quite suitable for current conduction, at 
higher $V_{SD}$ where occupation probability of major conducting state 
7e-1st-es reduces and that of 7e-gs increases, resulting in a stepwise 
behavior. Thus, there is no NDC in the I-V characteristics in the presence of 
higher magnetic field. 

Note that, generally it has been believed that magnetic flux needed to 
affect the molecular conduction is of the order of a magnetic flux quanta 
and required magnetic field is unrealistically huge in magnitude. However, 
as the bias dependent occupation probabilities and charge densities of 
transport active many-body states in our case are highly sensitive towards 
magnetic fields of much lower strength, we find evidence of modifications in 
I-V characteristic, even under a weak magnetic perturbation.

By analyzing the I-V characteristics of other molecule-electrode 
conformations, we find that magnetic field has very little effect on
the I-V characteristics in these cases. 

\begin{figure}
\centering
\includegraphics[width=0.8\textwidth]{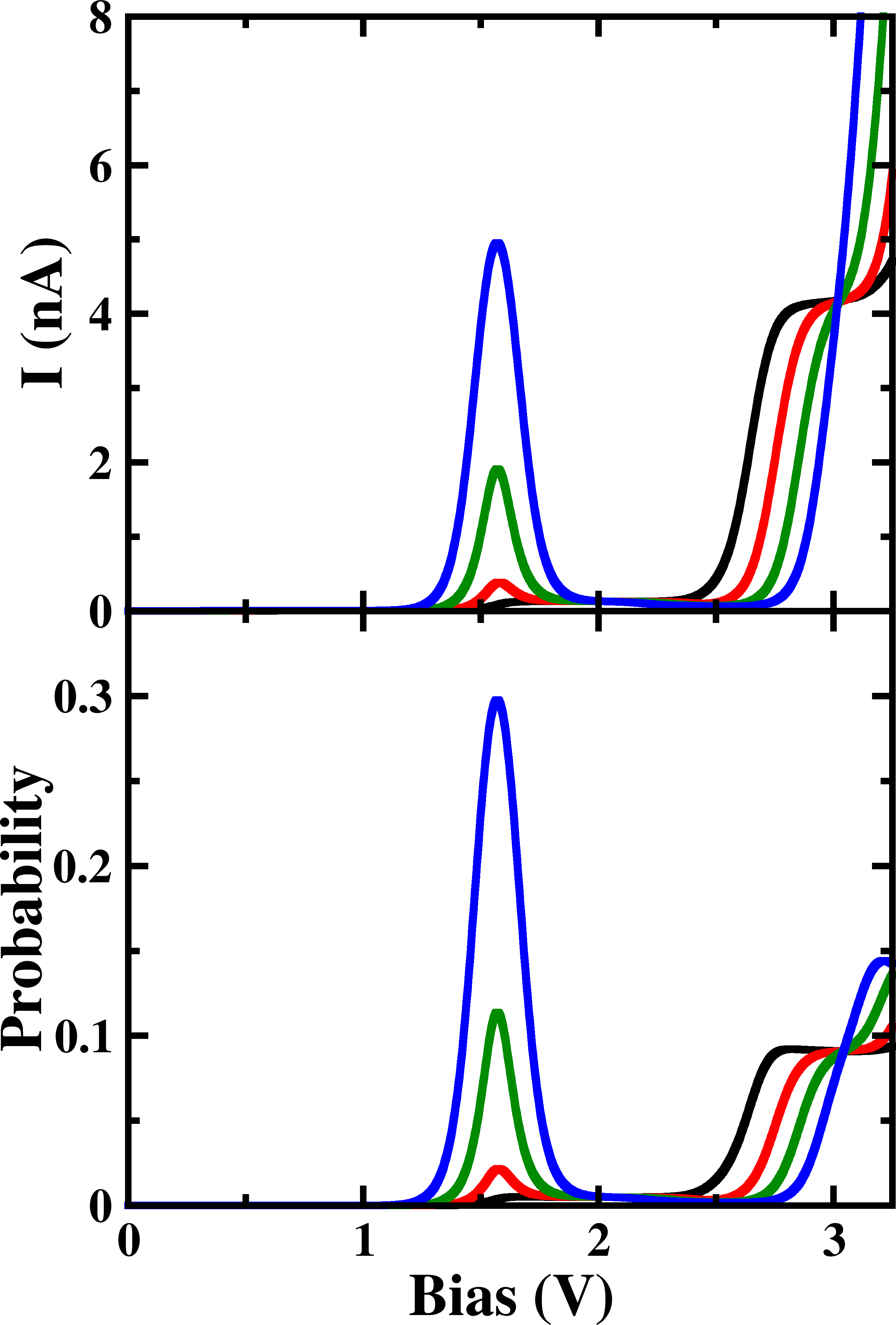}
\caption{ \label{gate1} Upper panel shows the I-V characteristic of boron-electrode connected junction at different gate voltage. 
Lower panel depicts the occupation probability of 7e-1st-es at $0.0 V < V_{SD}<3.5V$ where different gate voltage has been applied. 
Black, red, green and blue solid lines represent the current (upper panel) and probability (lower panel) with applied gate voltage of 0.15V, 0.10V, 0.05V and 0V, respectively.}
\end{figure}

Interestingly, other perturbing factor, i.e. applied gate voltage, can also 
significantly affect the resulting current. Particularly, the molecular 
junction where prominent NDC character appears in the absence of gate voltage, 
exhibits significant alteration of net current flow upon the application of 
gate-voltage. As can be seen in Fig. \ref{gate1}, negative differential 
conducting nature can efficiently be switched off and on by varying the 
strength of the applied gate voltage. Note that, strength of the gate voltage, 
used in the present study, has been regularly achieved in experiments. 
Looking at the probability distribution, it becomes evident that as the
strength of gate voltage increases, the probability of conducting 
7e-1st-es around $1.0 V < V_{SD}<2.0V$ gets diminished. Consequently, 
the NDC peak also disappears. Transition rates among the transport active 
many-body states (6e-gs to 7e-1st-es in present system) as well as their 
relative energies compared the electrodes fundamentally causes these kind of 
alteration in the I-V characteristics. 
   
\section{Conclusions}
To conclude, we have discussed the transport characteristics of cyclic 
molecular systems in sequential tunneling regime, applying well known kinetic 
equation approach. We chose our system to be a six-site molecule where donor, 
bridge and acceptor get arranged in a cyclic manner. Results clearly indicate 
that I-V character of presently investigated molecule strongly depend on the 
atomic nature of the sites, which are connected to the electrodes. When two 
different chemical species form the electrode-molecule coupling, prominent 
rectification of current appears. Importantly, when both B atoms are used as 
connecting sites, we find appearance of NDC in the device. All these exciting 
non-linear properties emerge due to the $V_{SD}$ dependent population and 
depopulation of low-lying excited states, which are quite different in terms 
of charge distribution pattern and hence nature of electron conduction. 
Fundamentally, the unequal charge distributions at different-sites of the 
molecule for these transport active many-body states determine the I-V 
characteristic. Particularly, the charge densities at the connecting sites 
to the electrode mainly control the current conduction through the molecular 
junction. Further, we also demonstrate the modulation of current conduction 
by applying perpendicular magnetic field. Depending on the strength, we find 
a number of interesting features, arising in the I-V characteristics. 
Thus, in this paper, we have successfully demonstrated various non-linear 
transport behavior appear for an intrinsically asymmetric heteronuclear 
cyclic molecule containing nano-junction.

\bibliography{many_body.bib}

\end{document}